\documentclass[aps,prl,superscriptaddress,twocolumn,floatfix,a4paper]{revtex4}

\usepackage{graphicx,graphics,epsfig}   
\usepackage{dcolumn}    
\usepackage{bm}         
\usepackage{amsmath}    
\usepackage{verbatim}   
\usepackage{color}      
\usepackage{subfigure}  
\usepackage{natbib,times}
\usepackage{amsmath,amsfonts,amssymb,graphics,graphicx,epsfig,color,natbib}

\newcommand{\PPR}[0]{P^{\text{PR}}}

\newcommand{\Pc}[0]{P^{\text{c}}}

\newcommand{\Pf}[0]{P^{\text{f}}}

\newcommand{\be}{\begin{equation}}
\newcommand{\ee}{\end{equation}}
\newcommand{\ba}{\begin{eqnarray}}
\newcommand{\ea}{\end{eqnarray}}
\newcommand{\ban}{\begin{eqnarray*}}
\newcommand{\ean}{\end{eqnarray*}}

\begin{document}

\title{Bound non-locality and activation}

\author{Nicolas Brunner}%
\affiliation{H.H. Wills Physics Laboratory, University of Bristol, Tyndall Avenue, Bristol, BS8 1TL, United Kingdom}
\author{Daniel Cavalcanti}
\affiliation{Centre for Quantum Technologies,
National University of Singapore, 3 Science Drive 2, 117543 Singapore,
Singapore}
\author{Alejo Salles}
\affiliation{Niels Bohr Institute, Blegdamsvej 17, 2100 Copenhagen, Denmark}
\author{Paul Skrzypczyk}
\affiliation{H.H. Wills Physics Laboratory, University of Bristol, Tyndall Avenue, Bristol, BS8 1TL, United Kingdom}

\date{\today}

\begin{abstract}
We investigate non-locality distillation using measures of non-locality based on the Elitzur-Popescu-Rohrlich decomposition. For a certain number of copies of a given non-local correlation, we define two quantities of interest: (i) the non-local cost, and (ii) the distillable non-locality. We find that there exist correlations whose distillable non-locality is strictly smaller than their non-local cost. Thus non-locality displays a form of irreversibility which we term bound non-locality. Finally we show that non-local distillability can be activated.
\end{abstract}

\maketitle

Entanglement and non-locality are both powerful resources for information processing \cite{RMP_horo,ekert91}. While entanglement has always been at the heart of quantum information science, there has recently been a growing interest in investigating non-locality from an information-theoretic perspective. On the one hand, quantum non-locality allows for the reduction of communication complexity \cite{CC}, as well as for information processing in the device-independent setting \cite{diqkd,randomness,stateest}, where one wants to achieve an information task without any assumption on the devices used in the protocol. On the other hand, in trying to understand why quantum non-locality is limited \cite{PR}, that is, why non-signaling correlations stronger than those allowed in quantum mechanics do not appear to exist in nature, it has been realized that strong non-locality enables a dramatic increase in information-theoretic power compared to the quantum case. For instance, certain post-quantum correlations collapse communication complexity \cite{vanDam,brassard,BS}, violate information causality \cite{IC} and macroscopic locality \cite{miguel}, and outperform quantum correlations for non-local games \cite{noah,GYNI}.

In general, in order to harness the information-theoretic power offered by a given type of resource, it is essential to understand how to quantify it. While this issue is rather well developed in the case of quantum entanglement \cite{RMP_horo}, much less is known for non-locality. We lack an adequate theoretical framework for tackling this problem, thus it is still not clear today what is a good measure of non-locality, and under which conditions two non-local correlations can be considered as equivalent.

The first tentative measures of non-locality were proposed in the context of Bell type experiments. For instance, one can consider the amount of violation of a Bell inequality or the resistance to noise---or to detector inefficiency---of a given set of correlations. However, 
it may happen that a set of correlations does not violate a given Bell inequality even if it is non-local. Moreover, it was recently shown that non-locality can be distilled \cite{foster}, that is by locally processing several copies of certain non-local correlations one can increase the amount of violation of a Bell inequality. Therefore, from an information-theoretic perspective, one needs better adapted measures of non-locality.


In the present paper, we study measures of non-locality based on the Elitzur-Popescu-Rohrlich (EPR2) \cite{EPR2} decomposition in the context of non-locality distillation. The idea of EPR2 consists of decomposing a given correlation $P$ into a purely local and a purely non-local part. The weight of the non-local part, minimized over all possible decompositions, then characterizes the non-locality of $P$. The EPR2 decomposition provides a natural framework for studying non-locality distillation. Given $N$ copies of $P$, corresponding to the non-local correlation $P^{ \times N}$, we identify two relevant quantities. The first one is the \emph{distillable non-locality}, which quantifies the amount of non-locality that can be extracted from $P^{\times N}$. The second is the \emph{non-local cost}, which quantifies how much non-locality is required in order to build $P^{\times N}$. We investigate these two quantities for a specific class of non-local correlations in the two-copy scenario. Interestingly we uncover a form of irreversibility, in that for certain undistillable correlations the non-local cost is strictly larger than the distillable non-locality. We term this effect \emph{bound non-locality} in analogy to bound entanglement \cite{BE}. We also provide an example of bound non-locality in the asymptotic limit ($N\rightarrow \infty$). Finally we demonstrate activation of non-local distillability, whereby an undistillable correlation can enhance non-locality distillation.

\emph{Measures of non-locality.} We consider the following scenario. Two remote parties, Alice and Bob, share a supply of non-local correlations, which from now on we shall refer to as non-local boxes. Formally these boxes are represented by a joint probability distribution $P(ab|xy)$, where $x,y$ denote the inputs of Alice and Bob respectively, and $a,b$ their outputs---here all alphabets are unspecified.

We consider a measure of non-locality based on the the EPR2 decomposition \cite{EPR2}, which consists in decomposing a non-local box $P$ into a convex mixture of a local part and a non-local part, that is
\ba\label{epr2} P(ab|xy) = (1-p_{\textsc{nl}})P_{\textsc{l}}(ab|xy)+ p_{\textsc{nl}} P_{\textsc{nl}}(ab|xy), \ea
where $P_{\textsc{l}}(ab|xy)$ is a local probability distribution and $P_{\textsc{nl}}(ab|xy)$ is a no-signaling probability distribution. The non-locality of the box $P$, denoted $C(P)$, is then obtained by minimizing the weight of the non-local part over all possible decompositions of the form \eqref{epr2}, i.e.
\ba\label{cost} C(P)= \min_{\text{decompositions}} p_{\textsc{nl}}. \ea
It follows that, for the optimal decomposition, the non-local part $P_{\textsc{nl}}(ab|xy)$ 
has unit weight, i.e. $C(P_{\textsc{nl}})=1$. $C(P)$ can be interpreted as the non-local cost of the box $P$, in the sense that this quantity represents the minimum amount of non-local resources required in order to construct $P$ 
\footnote{Note that here the non-local cost is defined in the single copy scenario.}.

An important property of $C(P)$ is that it cannot on average increase under local operations (LO), and is thus a meaningful measure of non-locality. Here local operations include relabelling of inputs and outputs. Note that in entanglement theory, the class of operations under which entanglement cannot increase is LOCC, where CC stands for classical communication. In the case of non-locality, communication between the parties is however not allowed, since it is a non-local resource.

A defining property of the measure $C(P)$ is that all non-local resources are treated on an equal footing. More precisely, all extremal non-local boxes are counted as equally non-local. As such we do not require any knowledge of the non-local part, namely which extremal boxes it involves. Since characterizing extremal non-local boxes is a hard problem, this property appears very advantageous. Moreover, it turns out that $C(P)$ can be computed efficiently using a linear program \cite{fitzi} (see below).

In the present paper we deal with the scenario where Alice and Bob share $N$ copies of a given box $P$. Formally they share the probability distribution
\ba P^{\times N}(\textbf{a} \textbf{b}|\textbf{x} \textbf{y}) = P(a_1b_1|x_1y_1) \times ... \times P(a_N b_N|x_N y_N) \ea
where we use the vector notation $\textbf{a} = \{a_1,...,a_N \}$ for the string of Alice's outputs and similarly for $\textbf{b},\textbf{x}$ and $\textbf{y}$.

First we would like to quantify the non-local cost of $P^{\times N}$. From the structure of $P^{\times N}$, it is easy to see that the weight of the non-local part is less than or equal to $1-(1-C(P))^N$, which is simply one minus the weight of the fully local part of the $N$ boxes. However, this represents only one possible decomposition of the form \eqref{epr2}, and we are by no means guaranteed that it is optimal. Indeed, instead of being given $N$ identical copies of $P$, we might hold a box behaving exactly as $P^{\times N}$ (i.e. represented by exactly the same probability distribution), but made using less non-local resources. Therefore the correct measure of the non-local cost of $P^{\times N}$ is given by its EPR2 decomposition, i.e. $C(P^{\times N})$.

In the context of Clauser-Horne-Shimony-Holt (CHSH) \cite{chsh}, when $x,y,a,b$ are all bits, it was recently shown that non-locality can be distilled \cite{foster,BS}. More precisely, by LO, which now include wiring together several copies of a box $P$, it is possible to obtain a box $P'$ which contains more non-locality. Note that in Refs \cite{foster,BS} the non-locality of a box was measured via its CHSH value, which, in this case, coincides with the measure of non-locality we adopt here.

It therefore appears natural to define the $N$-copy distillable non-locality of a box $P$, to be given by the maximal non-locality obtainable by wiring $N$ copies of $P$, that is
\ba\label{distillable}
D(P^{\times N}) = \max_{W} C(W[P^{\times N}])
\ea
where $W$ is an $N\rightarrow 1$ wiring, that is $W$ maps $P^{\times N}$ to a box $P'=W[P^{\times N}]$ which features inputs and outputs of the same size as the initial box $P$. A box $P$ is said to be $N$-copy distillable when $D(P^{\times N})>C(P)$.

Now that we are in a position to quantify how much non-locality can be extracted from $N$ copies of a box $P$, it is natural to compare this quantity to $C(P^{\times N})$, the non-local cost of $P^{\times N}$. A first simple observation is that
\ba\label{ineqs}
C(P) \leq D(P^{\times N}) \leq C(P^{\times N}) .
\ea
The left inequality follows from the fact that it is always possible for Alice and Bob to apply a trivial wiring, which consists in using only a single box and throwing away the $N-1$ remaining copies. The right inequality expresses the fact that it is impossible to extract more non-locality from a box than the amount of non-locality actually contained in the box. Importantly, the inequalities \eqref{ineqs} naturally link the EPR2 decomposition with non-locality distillation: $C(P)<C(P^{\times N})$ is a necessary condition for distillation to be possible.

A natural issue to investigate is reversibility. Can the non-locality contained in $N$ copies of a box $P$ always be extracted via distillation? In other words is $D(P^{\times N}) = C(P^{\times N})$? In the following we will answer this question in the negative. Furthermore, we will discuss an even stronger form of irreversibility. There exist boxes $P$ which cannot be distilled, although $N$ copies of $P$ contain a strictly larger amount of non-locality than a single copy. Formally this means $C(P) = D(P^{\times N}) < C(P^{\times N})$. We term this phenomenon bound non-locality, in analogy to bound entanglement. Below we provide examples of bound non-local boxes in the 2-copy setting, as well as an example in the asymptotic limit.




\emph{Bound non-locality.} From now on we focus on the CHSH scenario, where $x,y,a,b \in \{0,1\}$. We consider a 2-dimensional section of the no-signaling polytope \cite{barrett}, characterized as follows:
\ba\label{family} P(\xi,\gamma) \equiv \xi \PPR + \gamma \Pc + (1- \xi - \gamma )  \Pf \ea
with $\xi,\gamma \geq 0$ and $\xi + \gamma \leq 1$. Here we have used the following probability distributions: $\PPR (ab|xy) = \tfrac{1}{4}\left( 1 + (-1)^{a\oplus b \oplus xy} \right)$ is the Popescu-Rohrlich (PR) box, where $\oplus$ is addition modulo 2; $\Pc (ab|xy) = \tfrac{1}{4}\left( 1 + (-1)^{a\oplus b } \right)$ is a local box featuring unbiased but perfectly correlated outputs, $\Pf (ab|xy) = \tfrac{1}{8}\left( 2 + (-1)^{a\oplus b \oplus xy} \right)$ is the isotropic local box sitting on the CHSH facet below the PR box (i.e. a convex mixture of the PR box and white noise). The non-local cost of the boxes \eqref{family} is $C(P(\xi,\gamma))= \xi$, which follows from the fact that the PR box is the only extremal non-local box above the CHSH facet, on which both $\Pc$ and $\Pf$ lie. Thus \eqref{family} is the optimal EPR2 decomposition of $P(\xi,\gamma)$. Two classes of boxes that will be important below are: (i) isotropic non-local boxes $P_{\text{ISO}}(\xi) \equiv P(\xi,0)$, (ii) correlated non-local boxes $P_{\text{NLC}}(\xi) \equiv P(\xi,1-\xi)$.

The distillability of the boxes \eqref{family} was recently investigated. On the one hand, it was shown that isotropic non-local boxes cannot be distilled in the case of two copies \cite{toni_pur}. On the other hand, Refs \cite{foster,BS,closure,hoyer} presented distillation protocols for correlated non-local boxes. Below these protocols are referred to as follows: FWW for the protocol of \cite{foster}, BS for \cite{BS}, and ABLPSV for \cite{closure}.

We first focus on the case $N=2$, and characterize those boxes of the form \eqref{family} which can be distilled. It is possible to check, by an exhaustive search over all possible distillation protocols (see \cite{BS} for details), that the FWW and ABLPSV protocols are sufficient to characterize the distillable region. That is, $D(P(\xi,\gamma)^{\times 2})>C(P(\xi,\gamma))$ if and only if $P(\xi,\gamma)$ can be distilled via the FWW or ABLPSV protocol.

Next we compute $C(P(\xi,\gamma)^{\times 2})$, the non-local cost of two copies of $P(\xi,\gamma)$. This is done via a linear program \cite{fitzi}, which maximizes the weight of the local part, i.e. the quantity $p_{\textsc{L}}=1-p_{\textsc{NL}}$. The maximal local part, $p_{\textsc{L}}^*$, is the optimal value of the linear program
\ba  \text{max} & & \sum_{j=1}^{n}q_j, \\\nonumber   \text{subject to} & & \sum_{j=1}^{n}q_j D_j(\textbf{a} \textbf{b}|\textbf{x} \textbf{y}) \leq P^{\times 2}(\textbf{a} \textbf{b}|\textbf{x} \textbf{y}) \, , \,\,  q_j\geq 0  \ea
where the first condition should be understood as a vector inequality. Here $D_j(\textbf{a} \textbf{b}|\textbf{x} \textbf{y})$ denote the $n=4^8$ deterministic local strategies for the case of four inputs and four outputs. The non-local cost is then given by $C(P(\xi,\gamma)^{\times 2})=1-p_{\textsc{L}}^*$.

\begin{figure}[t]
\includegraphics[width=0.8\columnwidth]{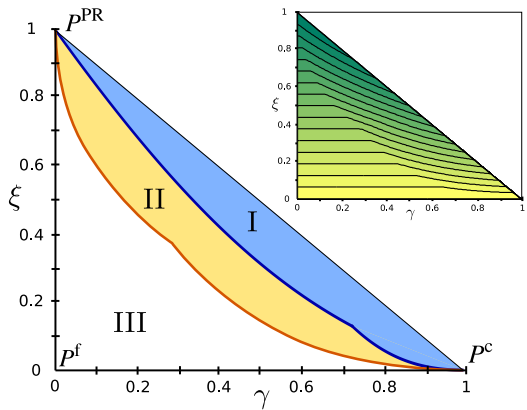} \caption{Section of the no-signaling polytope given by non-local boxes $P(\xi,\gamma)$. The region features three qualitatively different types of boxes: (I) 2-copy distillable boxes, (II) 2-copy bound non-local boxes, and (III) HH boxes. Inset: contour plot of the non-local cost of two copies, showing lines of constant $C(P^{\times 2}(\xi,\gamma))$.}\label{region}
\end{figure}

The results are presented in Fig.~\ref{region}. It shows three regions of qualitatively different kinds of boxes. The first region (I) corresponds to 2-distillable boxes, for which $C(P)<D(P^{\times 2})$. The second region (II) corresponds to 2-copy bound non-local boxes, for which $C(P)= D(P^{\times 2}) < C(P^{\times 2})$. The third region (III) corresponds to boxes such that $ C(P^{\times 2})=C(P)$; we term these `HH boxes'.

Indeed it would be interesting to investigate the case of more copies. Unfortunately very little is known beyond the case $N=2$. Still, a notable result is due to Fitzi et al. \cite{fitzi}, who investigated the scaling of the non-local part of $N$ copies of isotropic boxes as $N$ increases. Interestingly they found that $C(P_{\text{ISO}}(\xi)^{\times N})>C(P_{\text{ISO}}(\xi))$ for $N\geq3$ (see also \cite{dejan}). At first sight this result seemed to suggest that isotropic boxes could be distilled by a protocol involving three copies or more. However, our present findings, in particular the existence of bound non-locality, diminishes this hope. Another consequence of the result of \cite{fitzi}, is that HH boxes only exist in the case $N=2$. Thus this behaviour is a finite size effect.


Finally it is worth discussing the asymptotic case. A question of particular interest is whether bound non-locality can survive in the limit $N\rightarrow\infty$. Remarkably, the answer to this question is yes. An example is the isotropic box reaching the Tsirelson bound of quantum non-locality (CHSH$= 2\sqrt{2}$), i.e. the box $P_{\text{ISO}}(\xi)$ with $\xi=\sqrt{2}-1$. Clearly this box cannot be distilled, even with infinitely many copies, since the set of quantum correlations is closed under wirings \cite{closure}. Nevertheless, the result of \cite{fitzi} shows that for $N\geq3$, the non-local cost of $N$ copies exceeds the cost of a single copy. Therefore, this box is bound non-local in the asymptotic limit.


\emph{Activation of non-local distillability.} Above we have shown the existence of restricted forms of non-locality, such as bound non-locality. A natural question which arises now is whether the non-locality contained in such boxes can be activated.

Below we show that non-local distillability can be activated. First, we show that for any box $P_1$, there exists another box $P_2$ such that $D(P_1 \times P_2)> \max (C(P_1),C(P_2))$. In words, by combining one copy of $P_1$ and $P_2$, it becomes possible to achieve a task which would be impossible with one copy of either $P_1$ or $P_2$. Second, we present a stronger form of activation: for any 2-copy undistillable box $P_1$ and integer $N\geq 1$, there exists a box $P_2$ such that $D(P_1 \times P_2)> \max(D(P_1^{\times 2}),D(P_2^{\times N}))$, thus showing activation of undistillable non-locality.

Let us consider the following example. We take $P_1=P_{\text{ISO}}(\xi)$, and $P_2=P_{\text{NLC}}(\xi')$. Next we apply the BS protocol to $P_1\times P_2$ and obtain the box $P'=W_{\textsc{BS}}(P_1 \times P_2)$, which has non-local cost
\ba C(P')= \xi + \xi'(1-\xi)/8.  \ea
For all $0<\xi'\leq \xi <1$, we have that $C(P')>\xi=\max(C(P_1),C(P_{2}))$. Thus we get activation of non-locality. Note that this result holds for any box $P_1$ with binary inputs and outputs, since any such box can be `twirled' via LO to an isotropic box featuring the same non-local cost \cite{NS}.

Next, we note that $D(P_{\text{NLC}}^{\times N}(\xi')) \leq C(P_{\text{NLC}}^{\times N}(\xi'))\leq 1-(1-\xi')^N$. Thus, in the case $0<1-(1-\xi')^N \leq \xi<1$, we obtain that $C(P')>\max(D(P_1^{\times 2}),D(P_2^{\times N}))$. In other words, by combining one copy of $P_1$ and $P_2$, we obtain a box $P'$, with $C(P')>\xi$, which would be impossible from two copies of $P_1$ or from $N$ copies of $P_2$.

Moreover, when $P_1$ reaches Tsirelson's bound, i.e. $\xi=\sqrt{2}-1$, it is actually bound non-local in the asymptotic limit. Remarkably, one copy of the box $P_2$ can activate the bound non-locality of the box $P_1$ regardless of the amount of non-locality of $P_2$. We note the similarity between this example and the original example of activation of bound entanglement \cite{BE_activation}. There it was shown that by taking one copy of an entangled state $\rho$ which cannot be distilled without collective operations, and sufficiently many copies of a bound entangled state $\sigma_{BE}$, activation occurs. That is, the fidelity of $\rho$ with a maximally entangled state can be made arbitrarily close to one. This is indeed impossible for one copy of $\rho$, or for arbitrarily many copies $\sigma_{BE}$.

\emph{Discussion and open questions.} We investigated non-locality distillation using measures of non-locality based on the EPR2 decomposition, and showed the existence of bound non-locality. In addition, we presented examples of activation of non-local distillability. These examples show that \emph{any non-local box is useful for non-locality distillation}, in the sense it can be used to boost the distillation process of other boxes.

Let us comment on some open questions. A first issue concerns the computation of the measure. While the non-local cost can be computed efficiently, via a linear program, we had to run an exhaustive search over distillation protocols in order to determine the distillable non-locality. It would be interesting to find out whether distillable non-locality can be computed more efficiently, or at least whether meaningful bounds can be derived in a simpler way. One possibility would be to get a better understanding of the structure of sets of correlations which are closed under wirings \cite{closure}. For instance, given a initial set of boxes $S$, how could one characterize the set of boxes which can be generated by wiring arbitrarily many copies of boxes in $S$? In other words, how could one find the smallest closed set containing $S$?

A further interesting problem concerns the asymptotic behaviour of our measures. We have shown the existence of bound non-locality for two copies and at first sight one may wonder whether this is a finite-size effect. This is however not the case, as we have shown an example where bound non-locality survives in the limit $N\rightarrow\infty$. There is however much work to be done in order to understand the asymptotic regime.

Finally, it would be interesting to see how our measure of non-locality relates to others. In particular, it was recently shown that the PR box can be considered as a unit of bipartite non-locality \cite{unit}, in the sense that all bipartite non-local boxes can be simulated arbitrarily well using only PR boxes \cite{forster2}. This suggests another natural measure of non-locality, namely the minimal number of PR boxes required to simulate any box. It is noteworthy that computing this measure requires detailed knowledge of the non-local part, which is arguably undesirable. However, it would be interesting to understand the properties of this measure and how they relate to the measures presented here.

\begin{acknowledgements} We thank S. Popescu and M. Wolf for insightful comments. DC and AS thank A. Ac\'in for hospitality at ICFO, where part of this work was done. We acknowledge financial support from the UK EPSRC, EU project Q-ESSENCE, STREP COQUIT under FET-Open grant number 233747, the National Research Foundation and the Ministry of Education of Singapore.
\end{acknowledgements}


\begin{thebibliography}{19}

\bibitem{RMP_horo}
R. Horodecki, P. Horodecki, M. Horodecki, and K. Horodecki, Rev. Mod. Phys. {\bf 81}, 865 (2009).

\bibitem{ekert91}
A.~K. Ekert, Phys. Rev. Lett. {\bf 67},  661  (1991).

\bibitem{CC}
H. Buhrman, R. Cleve, S. Massar, and R. de~Wolf, Rev. Mod. Phys. {\bf 82}, 665 (2010).

\bibitem{diqkd}
A. Acin, N. Brunner, N. Gisin, S. Massar, S. Pironio, and V. Scarani, Phys. Rev. Lett. {\bf 98},  230501  (2007).

\bibitem{randomness}
S. Pironio \emph{et al}, Nature {\bf 464}, 1021 (2010).

\bibitem{stateest}
C.-E. Bardyn, T. C. H. Liew, S. Massar, M. McKague, V. Scarani, Phys. Rev. A {\bf80}, 062327 (2009).

\bibitem{PR}
S. Popescu and D. Rohrlich, Found. Phys. {\bf 24},  379  (1994).

\bibitem{vanDam}
W. van Dam, quant-ph/0501159  (2005).

\bibitem{brassard}
G. Brassard, H. Buhrman, N. Linden, A.~A. Methot, A. Tapp, and F. Unger, Phys. Rev. Lett. {\bf 96},  250401  (2006).

\bibitem{BS}
N. Brunner and P. Skrzypczyk, Phys. Rev. Lett. {\bf 102},  160403  (2009).

\bibitem{IC}
M. Pawlowski, T. Paterek, D. Kaszlikowski, V. Scarani, A. Winter, and M. Zukowski, Nature {\bf 461},  1101  (2009).

\bibitem{miguel}
M. Navascues, H. Wunderlich, Proc. Roy. Soc. Lond. A {\bf466}, 881 (2009).

\bibitem{noah}
N. Linden, S. Popescu, A.~J. Short, and A. Winter, Phys. Rev. Lett. {\bf 99},
  180502  (2007).

\bibitem{GYNI} M. L. Almeida, J.-D. Bancal, N. Brunner, A. Acin, N. Gisin, and S. Pironio, Phys. Rev. Lett. {\bf 104}, 230404 (2010).

\bibitem{foster}
M. Forster, S. Winkler, and S. Wolf, Phys. Rev. Lett. {\bf 102},  120401 (2009).

\bibitem{EPR2}
A. Elitzur, S. Popescu, and D. Rohrlich, Phys. Lett. A {\bf 162},  25  (1992).

\bibitem{BE}
M. Horodecki, P. Horodecki, and R. Horodecki, Phys. Rev. Lett. {\bf 80}, 5239 (1998).

\bibitem{fitzi}
M. Fitzi, E. H\"anggi, V. Scarani, and S. Wolf, arXiv:0811.1649 (2008).

\bibitem{chsh}
J.~F. Clauser, M.~A. Horne, A. Shimony, and R.~A. Holt, Phys. Rev. Lett. {\bf
  23},  880  (1969).

\bibitem{barrett}
J. Barrett, N. Linden, S. Massar, S. Pironio, S. Popescu, and D. Roberts, Phys.
  Rev. A {\bf 71},  022101  (2005).

\bibitem{toni_pur}
A.~J. Short, Phys. Rev. Lett. {\bf 102},  180502  (2009).

\bibitem{closure}
J. Allcock, N. Brunner, N. Linden, S. Popescu, P. Skrzypczyk, and T. Vertesi, Phys. Rev. A {\bf 80}, 062107 (2009).

\bibitem{hoyer}
P. H{\o}yer and J. Rashid, arXiv:1009.1668.

\bibitem{dejan}
D.~D. Dukaric and S. Wolf, quant-ph/0808.3317 (2008).

\bibitem{BE_activation}
P. Horodecki, M. Horodecki, and R. Horodecki, Phys. Rev. Lett. {\bf 82}, 1056 (1999).

\bibitem{NS}
L. Masanes, A. Acin, and N. Gisin, Phys. Rev. A {\bf 73},  012112  (2006).

\bibitem{unit}
J. Barrett, and S. Pironio, Phys. Rev. Lett. {\bf 95}, 140401 (2005).

\bibitem{forster2}
M. Forster, and S. Wolf, Ninth International Conference on QCMC, 1110, 117, AIP (2009).



\end{thebibliography}
\end{document}